\documentclass[aps,prl,twocolumn,showpacs,preprintnumbers,amsmath,amssymb]{revtex4}
\usepackage{bm}
\usepackage{amsmath}

\usepackage{graphicx}
\usepackage{dcolumn}

\begin{document}
\title{Classical Bound for Mach-Zehnder Super-Resolution}

\author{I. Afek}
\author{O. Ambar}%
\author{Y. Silberberg}%

\affiliation{%
Department of Physics of Complex Systems, Weizmann Institute of
Science, Rehovot 76100, Israel }
\date{\today}



\begin{abstract}
The employment of path entangled multiphoton states enables measurement of phase with enhanced precision. It is common practice to demonstrate the unique properties of such quantum states by measuring super-resolving oscillations in the coincidence rate of a Mach-Zehnder interferometer. Similar oscillations, however, have also been demonstrated in various configurations using classical light only; making it unclear what, if any, are the classical limits of this phenomenon. Here we derive a classical bound for the visibility of super-resolving oscillations in a Mach-Zehnder. This provides an easy to apply, fundamental test of non-classicality. We apply this test to experimental multiphoton coincidence measurements obtained using photon number resolving detectors. Mach-Zehnder super-resolution is found to be a highly distinctive quantum effect.
 \end{abstract}


\maketitle

\emph{Inroduction.}---Experimental observations which exhibit uniquely quantum mechanical behavior lie at the heart of quantum optics. Such observations establish the way we diffrentiate between the quantum and classical worlds. Notable examples include Bell's inequalities \cite{PRL81Aspect}, photon anti-bunching \cite{PRL77Kimble} and sub-Poissonian number statistics \cite{PRL83Mandel}. Negativity of the Wigner function is also widely used to verify non-classicality \cite{PRA00Davidovich}. The derivation of additional tests involving non-classical behavior has motivated many studies \cite{PRL04WAKS,PRL91Franson,PRA09Rivas,PRA07Dowling,PRL08Vogel,PRA02Knight}.

Recently, there has been a surge of interest in path entangled two-mode states of the form $|N::0\rangle_{a,b} \equiv \frac{1}{\sqrt{2}}  \left( |N,0\rangle_{a,b} +|0,N\rangle_{a,b}\right)$ known as NOON states \cite{CONTEMPPHYS08DOWLING}. In such states, all $N$ photons acquire phase as a collective entity thus enabling phase super-sensitivity \cite{CONTEMPPHYS08DOWLING} and sub-wavelength lithography \cite{PRL00}. A number of experimental implementations of NOON states exist to date \cite{PRL01SHIH,NAT04Steinberg,OE09,Afekarxiv}. In these experiments the high NOON state fidelity was demonstrated by measuring a sinusoidally varying $N$ photon coincidence rate with a reduced period corresponding to a wavelength $\lambda/N$. Such behavior, coined 'phase super-resolution' \cite{CONTEMPPHYS08DOWLING} is generally accepted as a signature of NOON states. However, it has been shown that similar measurements exhibiting high visibility super-resolution can be obtained using classical light only. This has been demonstrated using multiport linear interferometers with coincidence counting \cite{PRL07} or in a scheme for nonlinear sub-Rayleigh lithography \cite{Bentley04,Boyd06}.

\begin{figure}
\includegraphics{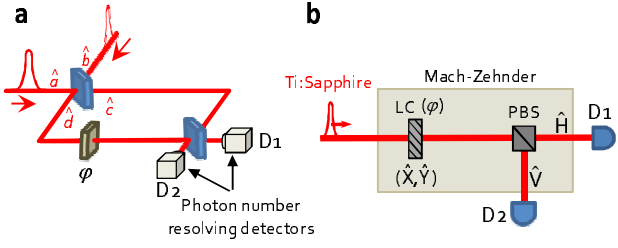}
\caption{ (color online) Experimental setup. (a) Schematic of the setup, a Mach-Zehnder interferometer with photon number resolving detectors at the output ports. (b) The actual setup used for obtaining the classical light results shown in Fig. \ref{fig:coinc_exp}. A horizontally (H) polarized pulsed Ti:Sapphire oscillator with $120$ fs pulses @$1$ MHz repetition rate (reduced from 80MHz using a pulse picker) is used as the light source. The MZI is implemented in a collinear inherently phase stable design using polarization. The phase, $\varphi$ is applied using a liquid crystal phase modulator at $45$ deg. ($X,Y$ polarizations). The photon number resolving detectors are Hamamatsu multi pixel photon counters (MPPC) module C10507-11-050U \cite{AfekPRA}. \label{fig:setup}      }
\end{figure}

It is therefore natural to ask whether super-resolution can be used at all as a direct, non-ambiguous test for non-classicality. Here we answer this question in the affirmative. We consider a Mach-Zehnder geometry similar to the one employed in the initial proposal for quantum lithography \cite{PRL00} and subsequently used in all NOON state generation experiments (see Fig. \ref{fig:setup}). We show that for any classical state with a positive-definite well behaved Glauber-Sudarshan $P$ representation \cite{PhysRev65Glauber}, there is a strict upper limit for the visibility of super-resolving fringes. The bound is derived analytically for arbitrary coincidences with $m,n$ photons in the two output ports respectively. We apply the bound to experimental multiphoton coincidences obtained using both classical and quantum light.  The classical results are found to approach the bound from below. For the quantum case, we employ NOON states obtained in an experiment we have recently reported  \cite{Afekarxiv} and a violation
of the classical bound with states of up to five photons is demonstrated.

\emph{Derivation.}---Consider a balanced Mach-Zehnder Interferometer (MZI) whose input modes  are denoted $a,b$ (see Fig. \ref{fig:setup}). The coincidence rate with $m,n$ photons in detectors $D_1,D_2$ respectively is denoted $C_{m,n}(\varphi)$ where the total number of photons is $N=n+m$. When employing a NOON state input, $C_{m,n}$ exhibits perfect sinusoidal fringes of the form $C_{m,n}(\varphi)\propto1+\cos(N\varphi-\delta)$ where $\delta=0,\pi$ depending on the values of $m,n$. This pattern exhibits $N$-fold phase super-resolution with $100\%$ visibility. Our goal is to derive a simple unambiguous criterion based on the visibility of $C_{m,n}$ for distinguishing between quantum and classical input states.
To this end, consider the Fourier series of an arbitrary coincidence function,
\begin{align}
C_{m,n}(\varphi) = \sum_{k=0}^{\infty} A_k \cos(k\varphi - \delta_k) .\label{eq:FT}
\end{align}
We are interested in the visibility of the $N$-fold oscillations, $|A_N/A_0|$, which will be referred to as the $N$-fold visibility. For pure sinusoidal oscillations with a constant background the $N$-fold visibility coincides with the conventional definition of visibility. Considering the $N$-fold visibility allows application of this bound to situations in which the measured coincidence has contributions oscillating at a number of different frequencies (see \cite{Afekarxiv} for example).
The $m,n$ coincidence rate for a MZI is given by \cite{PRL09PUENTES},
\begin{align}
C_{m,n}\left(\varphi\right)=\mathrm{Tr}\left[\hat{U}\left(\varphi\right)\hat{\rho}_{a,b} \hat{U}^{\dag}\left(\varphi\right) {\displaystyle \hat{\pi}_{n}^{1}\otimes\hat{\pi}_{m}^{2}}\right], \label{eq:traceequation}
\end{align}
where $\hat{U}\left(\varphi\right)$ is a unitary operator describing the MZI \cite{PRA86YURKE}, $\hat{\rho}_{a,b}=|\psi\rangle_{a,b} \langle\psi|$ is the input state density matrix and $\hat{\pi}_{n}^{1(2)}$ are the positive operator valued measures (POVM's) of detectors $D_{1(2)}$ \cite{PRL09PUENTES}. For the present derivation we consider ideal photon number resolving detectors implying $\hat{\pi}_{n}^{1}=\hat{\pi}_{n}^{2}=|n\rangle\langle n|$. First, we derive the $N$-fold visibility for an input state of the form $|\psi\rangle_{a,b}=|\alpha\rangle|0\rangle_{a,b}$. Here, $|\alpha\rangle$ is an arbitrary non-vacuum coherent state defined $|\alpha \rangle \equiv \sum_{n=0}^{\infty} e^{-\frac{1}{2}|\alpha|^2} \frac{\alpha^n}{\sqrt{n!}}|n\rangle$. Substitution in Eq. (\ref{eq:traceequation}) yields,
\begin{eqnarray}
\displaystyle C_{m,n}(\varphi) = \frac{e^{-|\alpha|^2}}{n!m!}|\alpha|^{2(m+n)}|\sin(\varphi/2)|^{2m}|\cos(\varphi/2)|^{2n}
\notag \label{eq:first}\\
= \frac{e^{-|\alpha|^2}}{n!m!}\left|\frac{\alpha}{2}\right|^{2(m+n)}|e^{\imath\frac{\varphi}{2}}-e^{-\imath\frac{\varphi}{2}}
|^{2m}|e^{\imath\frac{\varphi}{2}}+e^{-\imath\frac{\varphi}{2}}|^{2n}. \label{eq:second}
\end{eqnarray}
We denote the $N$-fold visibility for each photon number pair, $m,n$ by $\Gamma^{m,n}_{cl}$. Using simple combinatorial considerations we obtain
\begin{align}
\displaystyle \Gamma^{m,n}_{cl} = 2\left|\sum_{r=0}^{2n} (-1)^r {2n \choose r}{2m \choose n+m-r} \right|^{-1} .\label{eq:gamma_def}
\end{align}
We note that $\Gamma^{m,n}_{cl}$ does not depend on the specific choice of $\alpha$. In the following we prove that $\Gamma^{m,n}_{cl}$ is, in fact, the largest $N$-fold visibility obtainable for an arbitrary classical input state. We now consider an input state consisting of two arbitrary coherent states $|\psi\rangle_{a,b}=|\alpha\rangle|\beta\rangle_{a,b}$. We denote the $N$th fourier component of $C_{m,n}(\varphi)$ for this state by $A_N^{\alpha,\beta,m,n}$. Using some straightforward but somewhat lengthy algebra it can be shown that for such an input state the $N$-fold visibility is always bounded by $\Gamma^{m,n}_{cl}$ i.e.
\begin{eqnarray}
\left|A_N^{\alpha,\beta,m,n}/ A_0^{\alpha,\beta,m,n}\right|\leq\Gamma^{m,n}_{cl}.
\label{eq:bound1}
\end{eqnarray}
\begin{figure}[ht]
\includegraphics{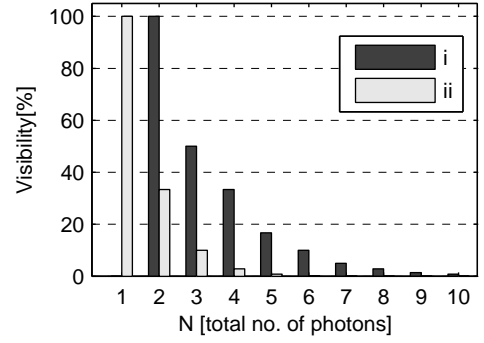}
\caption{\label{fig:vis_theory} The classical visibility bound, $\Gamma^{m,n}_{cl}$ (Eq. (\ref{eq:gamma_def})) as a function of $N=m+n$. For each $N$, the bound is shown for two choices of $m,n$: (i) $m,n=\lfloor N/2 \rfloor,\lceil N/2 \rceil$ i.e. the detection events are divided as equally as possible between $D_1$ and $D_2$ (see Fig. \ref{fig:setup}), (ii) $m,n=N,0$ i.e. all N photons are detected in $D_1$ and zero are detected in $D_2$. For a given $N$, the classically obtainable visibility for case (i) is always higher than for case (ii). However, for both cases $\Gamma^{m,n}_{cl}$ decays exponentially with $N$.  }
\end{figure}
%


\begin{figure}[htb]
\includegraphics{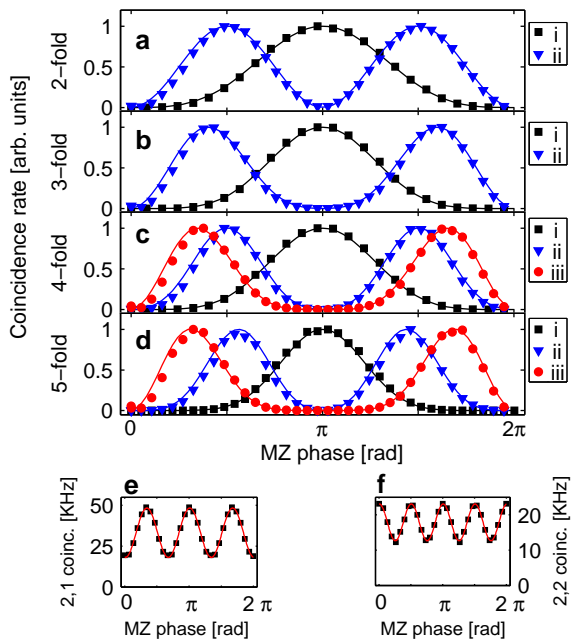}
\caption{\label{fig:coinc_exp} (color online) Experimental coincidence scans using classical light. (a-d) Coincidence rates (raw data, normalized) as a function of Mach-Zehnder phase with $m$ 'clicks' in $D_1$ and $n$ 'clicks' in $D_2$ for various values of $m,n$ (see Fig \ref{fig:setup}). The total number of photons $m+n$ is denoted by $N$. The input state is $|\psi\rangle_{a,b}=|\alpha\rangle|0\rangle_{a,b}$ where $|\alpha\rangle$ is a coherent state. The solid lines are obtained using an analytical model of the detector POVM's accounting for cross-talk, dark counts and losses \cite{AfekPRA}. Error bars are smaller than the displayed markers.  (a) $N=2$ (i) $m,n = 2,0$, (ii) $m,n = 1,1$. (b) $N=3$ (i) $m,n = 3,0$, (ii) $m,n = 2,1$. (c) $N=4$ (i) $m,n = 4,0$, (ii) $m,n = 2,2$, (iii) $m,n = 3,1$. (d) N=5 (i) $m,n = 5,0$, (ii) $m,n = 3,2$, (iii) $m,n = 4,1$. The classical visibilities in Fig. \ref{fig:vis_exp_theory}iii (blue bars) are obtained using these measurements. (e) Three-fold sinusoidal oscillations used for obtaining the experimental visibility of the $m,n=2,1$ classical case. To obtain the black squares, three copies of the raw data shown in panel bii were shifted by $0,\frac{2\pi}{3},\frac{4\pi}{3}$ and superimposed. The red line is the fitted Fourier series. The obtained 3-fold visibility is $45.22\%$. (f) Four-fold sinusoidal oscillations used for obtaining the experimental visibility of the $m,n=2,2$ classical case. Four copies of the  $m,n=2,2$ data shown in panel cii were shifted by $0,\frac{\pi}{2},\pi, \frac{3\pi}{2}$ and superimposed. The obtained 4-fold visibility is $29.19\%$.}
\end{figure}
Finally, we prove the bound holds for general classical inputs of the form $\hat{\rho}^{cl}_{a,b}=\int P(\alpha,\beta)_{a,b} |\alpha\rangle_{a}\langle\alpha| \otimes |\beta\rangle_{b}\langle\beta| d^2 \alpha d^2 \beta$  with a positive-definite, well behaved $P$ function $P(\alpha,\beta)_{a,b}$. These are precisely the quantum mechanical states which posses classical analogues \cite{PhysRev65Glauber}. Formally the $N$-fold visibility is a functional of $P$ which we denote $V_{m,n}[P(\alpha,\beta)]$. The bound ensues from the following inequalities,
\begin{eqnarray}
V_{m,n}[P(\alpha,\beta)] &=& \left| \frac{\int P(\alpha,\beta) A_{N}^{\alpha,\beta,m,n} d^2\alpha d^2\beta }{\int  P(\alpha,\beta) A_{0}^{\alpha,\beta,m,n}d^2\alpha d^2\beta} \right|\\
&\leq& \frac{\int P(\alpha,\beta)\left|\frac{A_{N}^{\alpha,\beta,m,n}}{A_{0}^{\alpha,\beta,m,n}}\right|A_{0}^{\alpha,\beta,m,n} d^2\alpha d^2\beta }{\int  P(\alpha,\beta) A_{0}^{\alpha,\beta,m,n}d^2\alpha d^2\beta}  \notag\\
&\leq& \frac{\int  P(\alpha,\beta) \Gamma^{m,n}_{cl} A_{0}^{\alpha,\beta,m,n}d^2\alpha d^2\beta}{\int P(\alpha,\beta)A_{0}^{\alpha,\beta,m,n}d^2\alpha d^2\beta}=\Gamma^{m,n}_{cl}\notag,
\label{eq:bound}
\end{eqnarray}
using the positive definiteness of $P$ and  Eq. (\ref{eq:bound1}). Note that $A_{0}^{\alpha,\beta,m,n}$ is positive by definition. This completes the proof and the main result of this letter.

The bound, $\Gamma^{m,n}_{cl}$, is plotted in Fig. \ref{fig:vis_theory} for a selection of $m,n$ pairs. In addition, the numerical values for $N\leq5$ are given in Table \ref{tab:hresult}. The function $\Gamma^{m,n}_{cl}$ decays exponentially with $N=m+n$. However, for a given $N$ the obtainable visibility is much higher when the photons are distributed equally (or almost equally for odd $N$) between the two output modes.  Interestingly, our result for pairs of the form $N,0$ coincide with those obtained by Bentley and Boyd \cite{Bentley04, Boyd06} in a proposal for sub-wavelength lithography using classical light thus showing their scheme reaches the classical bound.

We note that although the derivation uses a single spatial and spectral mode, its generalization to multiple modes is straightforward \cite{PhysRev65Glauber} and does not alter the final result. We have implicitly assumed the use of broadband detectors such as those typically employed for single photon detection. If one allows use of narrow-band multiphoton detection (such as atomic two photon absorption) and appropriate engineering of the input state then higher visibilities may be obtained \cite{OE04Peer,PRL06Scully}. The classical visibility limits of multiphoton interferences have been previously discussed \cite{PRA08agafonov,Klyshko93}. However, these analyses were limited to input states with randomly fluctuating phases and considered the standard rather than $N$-fold visibility which in our case can always be  $100\%$ (as seen experimentally in Fig. \ref{fig:coinc_exp}).  We point out that phase sensitivity beyond the standard quantum limit (known as phase super-sensitivity) is a distinct quantum effect and can be exploited for quantum noise reduction. However, it cannot be inferred directly from the visibility and requires additional knowledge about the efficiency \cite{PRL07,NJPH08}, making it much less straightforward as a test for non-classicality. The sensitivity criterion is also typically harder to beat experimentally than the present bound.

\begin{table}[t!] \caption{Classical visibility bound, $\Gamma^{m,n}_{cl}$, Eq. (\ref{eq:gamma_def}). Numerical values for $N\leq5$.} 
\centering     
\begin{tabular}{lccccccccccccccc}  
\hline    \\[-2.3ex]  \hline  \\ [-2.5ex]                      
$N$ &\multicolumn{2}{c}{2}& \multicolumn{2}{c}{3}& \multicolumn{3}{c}{4}& \multicolumn{3}{c}{5}&  \\ [0.5ex] 
\\ [-3.5ex] \hline \\ [-2.5ex]
$m,n\,$   &1,1&2,0&\,  2,1&3,0&\,  2,2&3,1&4,0& \, 3,2&4,1&5,0\\  
$\Gamma_{cl}^{m,n}(\%)$   &100&33.3& \;\!\!  50 & 10&  \; \!\! 33.3 & 20& 2.85& \;\!\!   16.67 &7.14 &0.79 \!\!\!\! \\
\hline                          
\end{tabular}
\label{tab:hresult}
\end{table}

\begin{figure}[ht!!]
\includegraphics{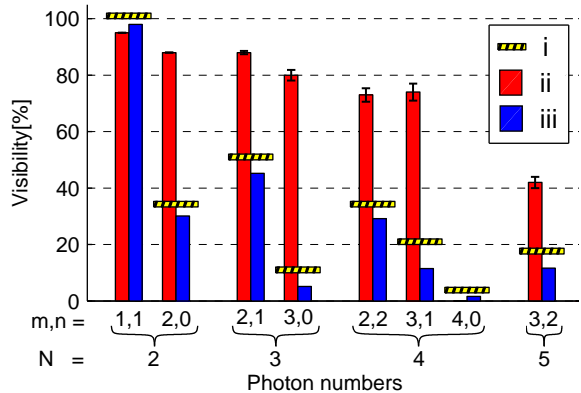}
\caption{\label{fig:vis_exp_theory} (color online) Comparison between the classical bound and experimental visibilities. Visibilities as a function of the number of coincidence 'clicks' in each of the MZI outputs. (i) The classical bound, $\Gamma^{m,n}_{cl}$, Eq. (\ref{eq:gamma_def}). (ii) Experimental visibilities obtained with NOON states using non-classical light \cite{Afekarxiv}. All the visibilities  violate the classical bound (except $4,0$ which was not measured and $1,1$ which cannot be surpassed). Error bars indicate $\pm \sigma$ statistical uncertainty. (iii) Experimental visibilities obtained using classical inputs (error bars are negligible). The experimental visibilities were obtained from the coincidence measurements in Fig. \ref{fig:coinc_exp}.  These visibilities are lower than the classical bound as expected, the discrepancy being mainly due to the cross-talk between detector pixels. }
\end{figure}

\emph{Experimental results.}---We now demonstrate the applicability of the bound using experiments done in both the classical and quantum regimes. In the classical case, the bound is approached from below by using a MZI with a coherent state input of the form $|\psi\rangle_{a,b}=|\alpha\rangle|0\rangle_{a,b}$. The MZI is implemented in a polarization based geometry as shown in Fig. \ref{fig:setup}b. We measured experimental multiphoton coincidences using two multi-pixel photon number resolving detectors \cite{AfekPRA}.  The coincidence sweeps are shown in Fig. \ref{fig:coinc_exp}a-d. For a given $N$ it is readily observable that the features are widest when the photons are all detected in one output. It is therefore not surprising that the visibility bound is lowest in this case. The $N$-fold visibility was obtained by fitting the experimental curves to a Fourier series of the form in Eq. (\ref{eq:FT}) truncated at $k=N$. For a given $N$, the fit was applied to $N$ copies of the data shifted by $0,2\pi\frac{1}{N},\ldots2\pi\frac{N-1}{N}$ and superimposed (see Fig. \ref{fig:coinc_exp}e-f). This eliminates the lower frequencies without affecting the $N$-fold visibility thus improving the accuracy of the numerical fit. Although in this case the slowly oscillating terms were eliminated in post-processing, it can be shown that for any value of $m,n$ there exists a classical input state which yields a coincidence rate of the form $C_{m,n}(\varphi)\propto 1+\Gamma^{m,n}_{cl}\cos(N\varphi+\delta)$ i.e. pure $N$-fold oscillations with a constant background. The experimentally obtained visibilities are shown in Fig. \ref{fig:vis_exp_theory}iii (blue bars). The measured visibility is lower than the classical bound for all $m,n$ as predicted. We note that limited overall efficiency doesn't affect the classically obtainable visibility since coherent states are transformed to coherent states by loss \cite{PRL06Bouwmeester}. Additional experimental imperfections such as cross-talk and dark-counts only serve to reduce the $N$-fold visibility.  Thus, the bound is expected to hold for raw experimental data and doesn't require compensation for detector imperfections.

Surpassing the bound requires use of non-classical light. To demonstrate this we use path-entangled multiphoton states with high NOON state fidelity that we have recently reported \cite{Afekarxiv}. Briefly, the experimental geometry was similar to the one used in Fig. \ref{fig:setup}, however one of the inputs was non-classical resulting in a highly entangled  state in modes $c,d$ of the MZI. The visibilities, obtained from raw data, are in violation of the classical bound. This verifies the quantum nature of the results and demonstrates the feasibility of surpassing the classical bound for a wide range of photon numbers in a realistic experimental setting, see Fig. \ref{fig:vis_exp_theory}ii (red bars).

\emph{Conclusion.}---We have derived a simple, practical test for non-classicality using MZI coincidence measurements and demonstrated its use.
 The quantity of interest is the visibility of $N$-fold oscillations in the $N$ photon coincidence rate. The classical bound on this quantity drops exponentially to zero with the total photon number $N$ while the quantum limit remains constant. This makes it an excellent indicator of non-classicality. It can be shown that only the NOON component of a given state in the MZI contributes to the $N$-fold oscillatory term. Thus, our results reflect the fact that the NOON state overlap resulting from any classical input state becomes exceedingly low as $N$ grows. We expect this result to serve as a benchmark in future experimental demonstrations of super-resolution.


\begin{acknowledgments}
Itai Afek gratefully acknowledges the support of the Ilan Ramon Fellowship.
Financial support of this research by the German Israel Foundation and the Minerva Foundation
is gratefully acknowledged. Correspondence and requests for materials
should be addressed to Afek. I ~(email: itai.afek@weizmann.ac.il).
\end{acknowledgments}

\end{document}